\documentclass[12pt]{nature}
\usepackage{times}
\usepackage[utf8]{inputenc}
\usepackage[numbers]{natbib}
\usepackage{gensymb}
\usepackage{graphicx}
\usepackage{lineno}
\usepackage{hyperref} 
\usepackage{multirow}
\usepackage{braket}
\usepackage{amsmath}
\usepackage[T1]{fontenc}
\usepackage{bm}
\usepackage{xcolor}
\usepackage{textcomp}
\usepackage[normalem]{ulem}
\usepackage{times}
\usepackage{chemformula} %allows chemistry formula
\usepackage[labelsep=endash]{caption}
\title{Weyl-Superconductivity revealed by Edge Mode mediated Nonlocal Transport}
\author{Wenyao Liu$^{1*}$, Gabriel Natale$^{1*}$, Camron Farhang$^{2*}$, Michael Geiwitz$^{1}$, Kewen Huang$^{1}$, Qishuo Tan$^{3}$, Xingyao Guo$^{4}$, Mason Gray$^{1}$, Vincent Lamberti$^{1}$, Jazzmin Victorin$^{1}$, Huairuo Zhang$^{5,16}$, James L. Hart $^{7}$, Vsevolod Belosevich$^{1}$, Xi Ling$^{3,17,18}$, Qiong Ma$^{1}$, Wan Kyu Park$^{6}$, Kenji Watanabe$^{11}$, Takashi Taniguchi$^{12}$, Judy J. Cha$^{7}$, Albert V. Davydov$^{5}$, Kin Chung Fong$^{8,19,20}$, Ethan Arnault$^{9}$, Genda Gu$^{10}$, Rui-Xing Zhang$^{13,14}$, Enrico Rossi$^{15}$, Jing Xia$^{2\dagger}$, Kenneth S. Burch$^{1\dagger}$}

\begin{document}
\maketitle
\begin{affiliations}
\item{Department of Physics, Boston College, Chestnut Hill, MA 02467, USA}
\item{Department of Physics and Astronomy, University of California, Irvine, CA 92697, USA}
\item{Department of Chemistry, Boston University, Boston, MA 02215, USA}
\item{Department of Physics, Hong Kong University of Science and Technology, Clear Water Bay, Hong Kong, China}
\item{Materials Science and Engineering Division, National Institute of Standards and Technology, Gaithersburg, MD 20899, USA}
\item{National High Magnetic Field Laboratory, Florida State University, FL 32310, USA}
\item{Department of Materials Science and Engineering, Cornell University, Ithaca, USA}
\item{Department of Electrical and Computer Engineering, Northeastern University, Boston, MA 02115, USA}
\item{Department of Electrical Engineering and Computer Science, Massachusetts Institute of Technology, Cambridge, MA 02139, USA.}
\item{Condensed Matter Physics and Materials Science, Brookhaven National Laboratory (BNL), Upton, NY 11973, USA.}
\item{Research Center for Electronic and Optical Materials, National Institute for Materials Science, 1-1 Namiki, Tsukuba 305-0044, Japan}
\item{Research Center for Materials Nanoarchitectonics, National Institute for Materials Science, 1-1 Namiki, Tsukuba 305-0044, Japan.}
\item{Department of Physics and Astronomy, University of Tennessee, Knoxville, TN 37996, USA}
\item{Department of Materials Science and Engineering, University of Tennessee, Knoxville, Tennessee 37996, USA}
\item{Department of Physics, William \& Mary, Williamsburg, VA 23187, USA}
\item{Theiss Research, Inc., La Jolla, CA 92037, USA}
\item{Division of Materials Science and Engineering, Boston University, 15 St. Mary’s Street, Boston, MA, 02215, USA}
\item{The Photonics Center, Boston University, 8 St. Mary’s Street, Boston, MA, 02215, USA}
\item{Department of Physics, Northeastern University, Boston, MA 02115, USA}
\item{Quantum Materials and Sensing Institute, Northeastern University, Burlington, MA}\\
{$^{*}$These authors contributed equally to this work.}\\
{$^{\dagger}$To whom correspondence should be addressed; E-mail: xia.jing@uci.edu, ks.burch@bc.edu}
\end{affiliations}

% \date{} % Include the date command, but leave its argument blank.
%%%%%%%%%%%%%%%%% END OF PREAMBLE %%%%%%%%%%%%%%%%
\baselineskip24pt

\newpage
%\noindent\textbf{
\begin{abstract} 
Topological superconductivity (TSC) hosts exotic modes enabling error-free quantum computation and low-temperature spintronics.\cite{read2000paired,mackenzie2003superconductivity,fu2008superconducting,qi2011topological,kallin2012chiral,linder2015superconducting,cai2023superconductor} Despite preliminary evidence of edge modes\cite{amet2016supercurrent,lee2017inducing,zhao2020interference,choi2022emergent,hatefipour2022a}, unambiguous signatures remain undetected\cite{serban2010domain,he2014correlated,zhang2017quantum,ikegaya2019anomalous}. Here, we report the first observation of protected, non-local transport from the edge modes of the potential Weyl-superconductor \ch{FeTe_{0.55}Se_{0.45}}. Namely resonant charge injection, ballistic transport, and extraction via edge modes.\cite{he2014correlated,zhang2017quantum,ikegaya2019anomalous}  An anomalous conductance plateau emerges only when topological, superconducting, and magnetic phases coexist, with source-drain contacts coupled via the edge. Moving the drain to the bulk switches the non-local transport process to a local Andreev process, generating a zero-bias conductance peak (ZBCP). The edge mode's topological protection is confirmed by its insensitivity to external magnetic fields and increasing temperatures until the spontaneous magnetization is substantially suppressed. Our findings provide a new methodology to demonstrate TSC edge states in \ch{FeTe_{0.55}Se_{0.45}} via topologically protected non-local transport.\cite{nayak2008non,qi2011topological,lutchyn2010majorana,alicea2012new,mong2014universal,albrecht2016exponential}\end{abstract}
%}

\section*{Introduction} 
Long-pursued are the non-trivial boundary states that would provide strong evidence for topological superconductivity (TSC), and could enable fault-tolerant quantum computation.\cite{fu2008superconducting,lutchyn2010majorana,kallin2012chiral,nayak2008non,qi2011topological,alicea2012new,yazdani2023hunting,linder2015superconducting,cai2023superconductor} An promising approach is propagating edge states from a chiral topological superconductor (C-TSC).\cite{serban2010domain,he2014correlated,zhang2017quantum,ikegaya2019anomalous} While numerous candidate systems have been experimentally explored,\cite{amet2016supercurrent,lee2017inducing,zhao2020interference,choi2022emergent,hatefipour2022a,vignaud2023evidence,yi2024interface} however, proving a TSC state requires directly probing the phase of the order parameter or the non-trivial properties expected from the emergent modes. As such, the clear identification of C-TSC is a long-standing experimental challenge since their predicted local electrical signals can be misleading and result from non-topological sources\cite{kayyalha2020absence,sarma2021disorder,hui2015bulk}. An alternative is exploring the undetected non-local and topologically protected responses of the edge states that produce unambiguous signatures in transport experiments.\cite{serban2010domain,he2014correlated,zhang2017quantum,ikegaya2019anomalous} With this in mind, we focused on the magnetized topological superconductor \ch{FeTe_{0.55}Se_{0.45}} that could produce a Weyl-superconducting phase with edge states (Fig. 1a)\cite{meng2012weyl,sau2012topologically,wu2021topological} when it combines a non-trivial topological band structure\cite{wang2015topological,zhang2018observation,li2021electronic,Lohani.2019}, superconductivity, and time reversal symmetry breaking (TRSB)\cite{mclaughlin2021strong,farhang2023revealing,zaki2021time,roppongi2025topology}. 

Here, we detect these previously unobserved non-local and topologically-protected signatures to explore potential Weyl-superconducting state in the iron-based superconductor \ch{FeTe_{0.55}Se_{0.45}} [Fe(Te,Se)]. The existence of TSC edge states is revealed via the observation of perfect transmission (Fig. 1b), long-range, decoherence-free, and non-local transport from source to drain (Fig. 1a) that is required the mediation of such states.\cite{serban2010domain,he2014correlated,zhang2017quantum,ikegaya2019anomalous} Specifically, to achieve this observation, we exploit the cleavable edges and exfoliatable nature of \ch{FeTe_{0.55}Se_{0.45}} via a new method of proper etching and placing contacts on straight and continuous edges (determined by AFM- Fig. 1c and cross-section STEM - Fig. 1d), where the top oxide enables transport primarily into the side. When the electrodes primarily touch crystalline edges of \ch{FeTe_{0.55}Se_{0.45}}, a robust conductance plateau is observed (Fig. 1g), and its non-local nature is confirmed by its disappearance upon moving the current drain to the bulk (Fig. 2). The decoherence-free and topologically protected nature of the conductance plateau is further demonstrated by its reproduction across numerous devices and in multiple laboratories (Fig. S7, S8), along with its immunity to thermal smearing (Fig. 3) and external magnetic fields (Fig. 4), which is a strong distinction from trivial Andreev bound state\cite{daghero2012strong}. Finally, the necessity of all three ingredients is confirmed by the correlation between the plateau's temperature evolution and the magnetization (Fig. 3), along with its absence in topologically trivial Fe(Te,Se) samples with similar magnetic and superconducting properties (Fig. 1e). 

\section*{Topology and TRSB in the Fe(Te,Se) Superconductor}
In the iron-chalcogenide superconductor \ch{FeTe_{1-x}Se_{x}} [Fe(Te,Se)], the concurrence of superconductivity and topological electronic band structure has been revealed in samples with suitable Te/Se ratios (x $\approx$ 0.45, $\textit{T}_{c} \approx 14.5~K$) \cite{wang2015topological,zhang2018observation,li2021electronic,Lohani.2019}. Furthermore, multiple experiments revealed TRSB below $T_{c}$ in Fe(Te,Se)\cite{mclaughlin2021strong,farhang2023revealing,roppongi2025topology,matsuura2023two}, with matching temperature dependence from Sagnac Magneto-Optical Kerr Effect (SMOKE) and $\mu$SR experiments demonstrating the TRSB state in the bulk\cite{roppongi2025topology}. The emergence of spontaneous TRSB in \ch{FeTe_{1-x}Se_{x}} could produce a new TSC regime, namely Weyl superconductivity, where Chiral superconducting edge states emerge along its side surfaces.\cite{sau2012topologically,meng2012weyl,wu2021topological,hu2024dislocation}

To explore this possibility, we first performed 4-terminal transport and SMOKE experiments to ensure TRSB and superconductivity indeed coexist in our \ch{FeTe_{0.55}Se_{0.45}} single-crystalline flakes and devices. These measurements revealed a sharp superconducting transition at $\textit{T}_{c} \approx 14.2~K$ (Fig. 1e) and the onset of a global spontaneous magnetization at $T_{Kerr}\approx 10~K < T_{C}$, with a rapid enhancement below $T^{*}_{Kerr}\approx 5~K$ (Fig. 1f). Here, we note that the SMOKE temperature dependence on our \ch{FeTe_{0.55}Se_{0.45}} flakes and devices are highly consistent with the $\mu$SR experiment\cite{roppongi2025topology}. Therefore, our devices contain superconductivity, topological band structure, and bulk TRSB below $T_{Kerr}$, whose proper combination may produce a Weyl TSC with gapless edge states\cite{sau2012topologically,meng2012weyl,wu2021topological,hu2024dislocation} (see Fig. 1a and Section 2 in the Supplementary Information). 

\section*{Device Design}

Theoretically, the emergence of C-TSC edge states produces distinguishing signatures in the differential conductance signal ($G[V,T] = \frac{dI}{dV}$) in non-local transport experiments\cite{he2014correlated,zhang2017quantum,ikegaya2019anomalous}. Specifically, the propagating edge modes generate topologically protected transport channels, and the self-Hermitian property ($\Gamma_{e}=\Gamma_{h}$) enforced by the TSC suppresses normal reflections (Fig. 1b)\cite{serban2010domain,law2009majorana}. For these reasons, if the current source and drain touch the same edge states, these modes mediate resonant current injection and extraction (Fig. 1a) along with ballistic transport (Fig. 1b). Thus, the charge transport through the C-TSC edge states is an unusual non-local process distinct from normal Andreev reflection (AR)\cite{park2010strong} or processes possible with normal Andreev bound states\cite{daghero2012strong}. Notably, a signature conductance plateau should result from the C-TSC edge modes constant density of states and absence of decoherence\cite{sau2012topologically,zhang2017quantum} whose width reflects their bandwidth (proportional to the exchange gap \(\Delta_{FM}\)\cite{sau2012topologically}), rather than superconducting gap \(\Delta_{SC}\). Due to its decoherence and barrier-free (i.e., resonant) qualities, the magnitude of the conductance plateau signal should be proportional to the number of new transport channels from the edge states and is temperature, magnetic field, and voltage-independent as long as the system remains in the C-TSC state. \cite{serban2010domain,ikegaya2019anomalous,he2014correlated,law2009majorana} Such signals are also distinct from the non-local transport associated with crossed Andreev reflection or elastic co-tunneling processes in trivial superconductors \cite{russo2005experimental,cadden2006nonlocal,morten2006circuit} which are rapidly suppressed via heat, external magnetic fields, or by separating the source and drain farther than the superconducting coherence length $\xi$.\cite{russo2005experimental,cadden2006nonlocal} 

However, these features were not observed in previous experiments measuring transport into \ch{FeTe_{0.55}Se_{0.45}}.\cite{park2010strong,tang2019quasi,gray2019evidence} Instead, some experiments revealed an anomalous conductance peak around zero bias voltage that was attributed to helical hinge modes.\cite{gray2019evidence} Indeed, those experiments were performed before the TRSB was detected and did not realize the device conditions needed to observe the non-local response. Here, we make a crucial advance by engineering electrical contacts that avoid the possibility of mode splitting between the current source and drain due to accidental steps on rough edges ('Line2' in Fig. 1c). Specifically, we exploited the tetragonal structure of Fe(Te,Se) and employed Atomic Force Microscopy (AFM) to pre-select (see Methods in SI) flakes with sharp (i.e., straight and continuous) edges (e.g., 'Line1' in Fig. 1c). For all data presented in the main text, we employed exfoliated Fe(Te,Se) flakes that displayed these sharp edges. 

Another crucial element is avoiding primary contact to the top surface (i.e. bulk). Indeed, the oxidation of the thin flake Fe(Te,Se) is known to occur rapidly, even inside a glovebox \cite{zalic2019fete}. Thus, similar to the challenge of making contacts to 2D TMDC materials, here, one cannot use the standard approach to making edge contacts in graphene. To overcome these challenges, we employed our 'cleanroom in a glovebox' to develop a simple method to realize edge contact for thin flake Fe(Te,Se). Specifically, we optimized the time and power of the Ar plasma just before metal deposition. This maintained the top oxide while making Ohmic contacts to the cleaved edge (see cross-section STEM image in Fig. 1d) with normal state contact resistance $R_{N} < 100~\Omega$ (Additional details regarding plasma parameters and their effects see Methods.)

In addition, we prepared \ch{FeTe_{0.55}Se_{0.45}} devices with a thick hexagonal Boron-Nitride (hBN) that partially covers one edge and the surface (Fig. 2f), providing a control that allows us to contact only the bulk as well as the sharp edges. To evaluate the Te/Se ratio, the exfoliated edges crystallinity, and the quality of the contacts and crystals for the devices utilized in this work (see Methods in SI), we also performed Selected Area Electron Diffraction (inset Fig. 1c), Scanning Transmission Electron Microscopy (Fig. S1), and X-ray Energy Dispersive Spectroscopy (Fig. S2). 

\section*{Non-Local Response}

We now turn to this work's key observation,  the bias-independent conductance plateau observed in T-FTS\#1 (see Fig. 1g). This result is consistent with TSC edge states but distinct from previous point-contact measurements showing zero-bias anomalies\cite{gray2019evidence} or the characteristic double-peak structure of AR.\cite{park2010strong,tang2019quasi} In T-FTS\#1, the plateau feature only appears within $\pm 0.4$ mV, which is inside $\Delta_{SC}~(\approx 3~ meV)$\cite{park2010strong,tang2019quasi,wang2018evidence}, and has a far larger conductance enhancement than in \ch{FeTe_{0.4}Se_{0.6}} or other superconductors\cite{park2010strong,tang2019quasi}. Consistent with detecting edge states' non-local signal and the requirement of a continuous path of all edge modes connecting the source and drain, the plateau becomes a ZBCP if even a single step occurs between the contacts. Nonetheless, the plateau can be observed if the source and drain are connected to different sides of the flake, so long as they are attached to one continuous edge (see the sharp edge in Fig. 1c SAED and profile 'Line1').  

Next, we test the plateau's non-local origin, a key feature of the TSC edge states. The differential conductance is probed by applying a current and measuring the resulting voltage of the source with respect to the bulk superconductor via a third contact ($G_{ij,mn}=dI_{ij}/dV_{mn}$, with $i$, $j$ the current and $m,n$ the voltage leads). In the 3-terminal measurement (Fig. 2a-c), there is no voltage drop in the bulk of Fe(Te,Se). As such, the drain's location should not affect $G_{ij,mn}$, which typically results from the local AR process.\cite{park2010strong,tang2019quasi} In contrast, the non-local $G_{ij,mn}$ from edge states should be very sensitive to the drain location, as it involves coupling the source and drain through the edge states (Fig. 1a). 

To test this we measured $G_{ij,mn}$ in the double-edge-lead configuration (Fig. 2a),  where the current source (electrode\#1) and drain (electrode\#3) are strictly connected through the sharp edge ($G_{DE}[V,T]=G_{13,12}[V,T]$, see Fig. 2f). Next, we evaluated the single-edge-lead configuration (Fig. 2b), where the source is still connected to the edge, but the drain only touches the surface ($G_{SE}=G_{14,12}$, Fig. 2f). This completely forbids non-local transport but still allows local tunneling into the edge states. Lastly, we checked the surface-lead configuration (Fig. 2c) where the source lead is only in contact with the flake surface ($G_{Sur}=G_{41,42}$ or $G_{43,42}$, Fig. 2f). In the surface-lead configuration, we expect to only observe the standard AR process into a superconductor. 

Consistent with this hypothesis, upon switching from double-edge-lead to single-edge-lead configuration (Fig. 2d), the conductance plateau ($G_{DE}$) becomes a zero-bias conductance peak (ZBCP, $G_{SE}$). This drastic conductance change confirms that the transport process in \ch{FeTe_{0.55}Se_{0.45}} anomalously depends on the drain contact and, thus, its non-locality. Indeed, this is consistent with suspending non-local transport through the edge states by moving the drain to the bulk, forcing the injected electrons to leave the superconductor via Cooper pairs rather than resonantly through the edge states. Thus, the ZBCP results from the local AR process at the source contact. As discussed in previous works, the ZBCP is not entirely understood may be explained by several mechanisms consistent with the hypothesis of locally probed TSC edge states, such as Andreev edge states (AES)\cite{law2009majorana,he2014correlated,zhao2020interference}. 

We also found that the non-local conductance plateau response was successfully detected in several devices, e.g., T-FTS\#2, even when source and drain contacts are separated by $5~\micro m$ (Fig. 2f) or on different sides of the flake. This displays the long-range coherence of the edge states\cite{ikegaya2019anomalous}, which extends orders of magnitude beyond the \ch{FeTe_{0.55}Se_{0.45}} superconducting coherence length ($\xi\approx 3~nm$) \cite{galluzzi2019transport}. Nonetheless, we found the plateau and ZBCP require contacting the edge, as seen in measurements using the surface-lead configuration on the same T-FTS\#3 device (Fig. 2d,e). As expected, the corresponding $G_{Sur}$ curves shown in Fig. 2e exhibit the double-peak feature at $\pm 3~mV$, consistent with local AR into the $\Delta_{SC}$\cite{zhang2018observation} and previous point-contact experiments on \ch{FeTe_{0.55}Se_{0.45}}\cite{park2010strong,tang2019quasi}. This occurs despite the edge and top contacts possessing similar $R_{N}$ and proves that the plateau and ZBCP disappear when the source only touches the bulk.

The controlled switching from a plateau, ZBCP, or normal AR $G_{ij,mn}$ spectra by changing source-drain configuration in the same device (Fig. 2d,e) strongly points to the existence of edge states and not experimental artifacts. Nonetheless, a fully quantitative description of the plateau from C-TSC edge states exceeds the scope of previous single-mode models and requires further theoretical treatment\cite{serban2010domain,he2014correlated,ikegaya2019anomalous,casas2024long}. However, as shown in Fig. 2g, we found that the height of the conductance plateau is approximately proportional to the thickness of \ch{FeTe_{0.55}Se_{0.45}}. Such thickness dependence is consistent with edge states with multiple channels (inset Fig. 2g).

\section*{Requirement for Topological States}

To test if the observed non-local transport is of topological origin, we performed identical-configuration experiments but on devices from Fe(Te,Se) with topologically trivial band structure, while its superconducting $\textit{T}_{c}$ and magnetism are still similar to \ch{FeTe_{0.55}Se_{0.45}} (Fig. 1e and Fig. S14 in SI). Specifically, we examined three devices with reduced Te-concentration \ch{FeTe_{0.4}Se_{0.6}} (N-FTS\#1-3) and another made trivial by excess Fe (\ch{Fe_{1+y}Te_{0.55}Se_{0.45}})\cite{li2021electronic,farhang2023revealing}.  Consistent with previous works,\cite{park2010strong,tang2019quasi} all devices made from these topologically trivial Fe(Te,Se) exhibited a low temperature conductance far below the edge state and possessed a double-peak feature in all contact configurations. A specific example is shown in Fig. 1h for the $G_{DE}$ response from N-FTS\#1. In addition, the temperature and magnetic field dependence of $G_{DE}$ was consistent with local AR (See Fig. S4 in SI). This strongly suggests that observing the plateau requires a topologically non-trivial band structure. 

To further ensure the topological nature, we tested the robustness of the non-local transport response by determining the reproducibility of the conductance plateau. Fig. 2d shows the conductance plateau only in the double-edge-lead measurement on different \ch{FeTe_{0.55}Se_{0.45}} flakes (T-FTS\#2-\#3, Fig. 2f). Nearly identical non-local conductance plateaus have been observed from over ten \ch{FeTe_{0.55}Se_{0.45}} devices (Fig. S7), including one with Al rather than Cr/Au leads (Fig. S8A). Additionally, similar results were obtained from experiments using various excitation schemes (AC+DC and purely DC) in multiple laboratories (See Differential Conductance Measurements section in SI). 

We also note the non-locality of the response is incompatible with other possible origins for the plateau and ZBCP features (e.g., the Klein-paradox effect from Dirac surface states\cite{lee2019perfect} or non-spectroscopic effects, including Joule heating\cite{gifford2016zero}). Nonetheless, we further excluded these via additional experiments that confirmed the observed signals were truly spectroscopic. Specifically, additional 2-terminal measurements revealed similar $G_{ij,kl}(V,T)$ (Fig. S6F), while 4-terminal measurements always reveal a zero resistance state (Fig. S6E). In addition, we found the critical current in Fe(Te,Se) devices to be well above the bias current applied in this work (Fig. S6D). Lastly, as discussed below, the temperature and magnetic field dependence of the plateau signal is at odds with these other possible origins.

\section*{Temperature Evolution and Magnetic Order}

To ensure the plateau signal results from edge state-mediated non-local transport, we must confirm the topological protection. To this end, we compared the temperature evolution in all three cases: plateau ($G_{DE}$), ZBCP ($G_{SE}$), and normal AR ($G_{Sur}$). Specifically, in normal point contact measurements, the local AR process is sensitive to scattering and Fermi smearing. Thus, $G_{SE}$ or $G_{Sur}$ will be smoothly smeared and suppressed until $\textit{T}_{c}$\cite{park2010strong,tang2019quasi}. However, for a conductance plateau from non-local transport through edge states, the response is protected from thermal smearing unless magnetic or superconducting orders are suppressed\cite{wu2021topological,hu2024dislocation}. 

Starting with the temperature evolution of $G_{DE}$, as shown in Fig. 3a \& 3d, the plateau response can be divided into three regimes (see the white dashed lines). The differential conductance is temperature-independent in the first regime ($ T < 5~K$). Then, the plateau narrows after entering the second regime ($5~K <  T < 10~K$), and finally in the third regime ($T > 10~K$), the zero-bias conductance $G(0,T)$ rapidly diminishes and fully vanishes before reaching $\textit{T}_{c}$ = 14.2 K. Meanwhile, in the Blonder-Tinkham-Klapwijk (BTK) model that describes local AR  \cite{park2010strong,tang2019quasi} (see simulated results in the Supplementary Information - Fig. S3), if the plateau results from perfect AR with an energy scale ($\Delta\approx 0.4~meV$)\cite{lee2019perfect}, it would suffer thermal smearing even in the first regime. In contrast, despite the larger energy scale ($\approx 3~meV$), $G_{sur}(V,T)$ (Fig. 3b) is immediately thermally smeared with increasing temperature and fades out at $\textit{T}_{c}$, in good agreement with previous experiments and the BTK model\cite{park2010strong,tang2019quasi}. 

To understand the non-BTK temperature dependence of the plateau signal, it is useful to consider that the existence of the C-TSC edge states depends on the presence of the magnetic order. Thus, we performed SMOKE measurements to study the magnetization in our Fe(Te,Se) flakes and devices as the polar-Kerr signal (\(\theta_{K}\)) is proportional to the perpendicular component of the magnetization \(M_{\perp}\)\cite{argyres_theory_1955}. As displayed in Fig. 3c, we found that $\theta_{K}$ is identical regardless of the training field's magnitude, which rules out its origin from pinned vortices. Furthermore, the transition temperature was universal on numerous flakes under different conditions: as exfoliated, in devices, and with various thicknesses and protective layers (Fig. S10). Here, we observed two key temperatures for the magnetic order: its onset at $T_{kerr} = 10 K$ and rapid rise in the Kerr signal below $T^{*}_{kerr} = 5 K$. These temperature scales agree well with the temperature evolution of the plateau spectra (Fig. 3a \& 3d).

To further understand the correlation between magnetization and the transport signal, we compare the temperature dependence of \(\theta_{K}\) with the normalized $G(0,T)$ curves for the three cases explored in this work. As displayed in Fig. 3d, we found that the plateau and ZBCP are suppressed at 10 K, which is $T_{kerr}$ rather than $T_{c}$. Moreover, $G_{DE}(0,T)$ (plateau) displays the same three regimes as seen by SMOKE (Fig. 3d). Such a strong correlation between SMOKE and the non-local transport signals supports the realization of TRSB-induced C-TSC state with corresponding edge states in Fe(Te,Se)\cite{sau2012topologically,wu2021topological}. Nonetheless, we expect the edge states' bandwidth to be determined by the magnetization, not just the superconducting gap (as in normal AR). To test this hypothesis, as seen in Fig. 3e we compared the voltage width of the conductance plateau (extracted from Fig. 3a) to $\theta_{K}$. These have a striking correspondence: as the temperature increases to $T_{Kerr}^{*} = 5~K$ (regime one), \(\theta_{K}\propto M_{\perp}\) rapidly decreases consistent with re-orientation of the magnetic easy-axis (see Supplementary), while the plateau spectra remain constant. Upon entering regime two ($T_{Kerr}^{*} \leq T \leq T_{Kerr}$), where the rate of change of \(\theta_{K}\propto M_{\perp}\) is reduced, the plateau signal and width correspondingly shrink. Finally, in regime three, upon approaching $T_{Kerr}\approx10~K$, $G_{DE}$, $G_{SE}$ and $\theta_{K}$ disappear, well below the $\textit{T}_{C}\approx 14.2~K$.

\section*{External Magnetic Field}

As a final check, we applied an external magnetic field at 1.5~K. As explained later, this should not affect the non-local conductance signal from C-TSC modes as long as the magnetic and superconducting orders are preserved. As shown in Fig. 4a, the conductance plateau is completely unaffected by external magnetic fields up to 8 T, applied along all three crystallographic directions. This is consistent with non-local transport through edge states as the applied fields are much smaller than the upper critical field ($H_{c2}\approx40 - 80~T$ \cite{galluzzi2019transport}) and will not change the magnetic order (Fig. 3c). Nonetheless this response is in stark contrast to that expected from Andreev bound states or topological helical modes in a time-reversal-symmetric system, where the magnetic field immediately lifts the degeneracy or topological protection. In addition, for local AR processes, the spectra will be modified by magnetic fields applied out of the plane ($B_{\perp}$), which generate vortices or screening currents along the edges of the device (Doppler shift).\cite{tanaka2003doppler,Zareapour2017a}

To this end, we measured the local AR in our devices by focusing on the magnetic field dependence of $G_{SE}$ and $G_{Sur}$. As expected from the Doppler shift to local tunneling into edge modes, we observe the ZBCP is quickly suppressed and broadened by $B_{\perp}$ (see Fig. 4b). In contrast, we find an in-plane magnetic ($B_{||}$) field 
only mildly affected the ZBCP (Fig. 4b bottom panel and its inset), consistent with $B_{||}$  generating small screening currents due to slight misalignment\cite{tanaka2003doppler,Zareapour2017a}. Lastly, as shown in Fig. 4c,  $G_{Sur}$ reveals a magnetic-field-induced signal reduction within the superconducting gap's voltage scale ($\approx 3~mV$), consistent with normal local AR. Thus, the topological protection and non-locality of the plateau in $G_{DE}$ is further confirmed by its distinct response to magnetic fields from that of the local probes (i.e., $G_{SE}$ and $G_{Sur}$). 

%\\\\\\\\\\\\\\\\\\\\\\\\\\\\\\\\\\\\\\\\\\\\\\\\\\\\\\\\\\\\\\\\\\\\\\
\section*{Discussion and Outlook}

Our results provide the first unambiguous, long-sought evidence of C-TSC edge states in \ch{FeTe_{0.55}Se_{0.45}} via the measurement of non-local transport mediated by chiral edge states. The non-locality is confirmed by employing varying contact configurations; namely, the signal switches from a plateau to a zero-bias peak upon moving from the double-edge-lead to single-edge-lead configuration. This dependence on the source and drain location is consistent with the switch from non-local to local tunneling into a C-TSC edge mode,\cite{law2009majorana,zhao2020interference} while placing the source on the surface produces the typical AR conductance signal. The plateau signal's reproducibility across devices and laboratories, its observation when source and drain are separated by more than $10^{3}$ times the superconducting coherence length, and its resistance to thermal smearing or external magnetic fields confirm the non-local, coherent, and topologically protected nature of the chiral edge modes in \ch{FeTe_{0.55}Se_{0.45}}. In addition, the close correspondence of the plateau's temperature dependence with the TRSB signal and the requirement for a topologically non-trivial normal state point to its origin being from a C-TSC state.

Taken together, our results prove the existence of TSC edge states by revealing their non-local and topologically-protected nature. Nonetheless, further theoretical and experimental efforts are needed to establish the specific superconducting state (i.e., Weyl, p-wave) responsible for the chiral edge states in this single-material platform. Specifically, although most responses are in good agreement with theoretical works describing the non-local transport through C-TSC edge states\cite{he2014correlated,zhang2017quantum,ikegaya2019anomalous}, such as a resonant co-tunneling or crossed-Andreev process\cite{he2014correlated}, to directly compare with the experiments the current theoretical models must be extended to multiple edge-states or 3D C-TSC systems. Furthermore, to account for the large edge conductance, one might need to consider the multi-band superconductivity of Fe(Te,Se) that could produce a topological superconducting state with a Chern number larger than 1 \cite{mascot2022topological}. Further clarity is also needed on the role of strong correlations in generating the unique magnetic, superconducting, and topological states of \ch{FeTe_{0.55}Se_{0.45}}.\cite{Kim.OrbitalSelectiveFTS.2024} Alternatively, it will be highly desirable to employ quantum point contacts to measure TSC edge modes' non-Abelian statistics. Beyond the fundamental interest in this new regime, it will be crucial for Majorana circuits, where magnetic domains are used to manipulate the edge states. 

\bibliographystyle{naturemag}
\bibliography{Reference.bib}

\section*{Acknowledgments}
We thank Fan Zhang, Ilija Zeljkovic, Ziqiang Wang, and Jigang Wang for helpful discussions. Disclaimer by NIST: Certain commercial equipment, instruments, software, or materials are identified in this paper in order to specify the experimental procedure adequately; Such identifications are not intended to imply recommendation or endorsement by NIST, nor it is intended to imply that the materials or equipment identified are necessarily the best available for the purpose.
\noindent\textbf{Funding:} 
The National Science Foundation, Award No, supported the work of W.L., M.Gray, and J.V. MRI-2117711, DMR-200334 and DMR-2310895. M.Geiwitz and K.H. are grateful for the support of the Office of Naval Research under Award number N00014-23-1-2539. G.N. and V.L. are grateful for the support of the National Science Foundation (NSF) EPMD program via grant EPMD-2211334. The AFSOR, Grant FA9550-20-1-0282, supported the work of K.S.B.. Transport measurements at Boston College were enabled by equipment provided through AFOSR DURIP award FA9550-20-1-0246. The Gordon and Betty Moore Foundation EPiQS Initiative, Grant\#GBMF10276 and and NSF award DMR-2419425 supported the work at UC Irvine. The work at BNL was supported by the US Department of Energy, oﬃce of Basic Energy Sciences, contract no. DOE-sc0012704. Q.M. and V.B. acknowledge support from the NSF CAREER award DMR-2143426 as well as the Alfred P. Sloan Foundation. J.C. and J.H. acknowledge support from the Gordon and Betty Moore Foundation EPiQS Initiative, Grant\#GBMF9062.01. H.Z. acknowledges support from NIST Cooperative Agreement\#70NANB22H101. E.R. acknowledges support from the US Department of Energy, Office of Basic Energy Sciences, via Award DE-SC0022245. The NSF\/DMR-2003405, NSF\/DMR-1644779, and the State of Florida supported the work at NHMFL and FSU. K.W. and T.T. acknowledge support from the JSPS KAKENHI (Grant Numbers 21H05233 and 23H02052) , the CREST (JPMJCR24A5), JST and World Premier International Research Center Initiative (WPI), MEXT, Japan. Q.T. and X.L. acknowledge the support by the National Science Foundation (NSF) under Grant No. 1945364 and US Department of Energy (DOE), Office of Science, Basic Energy Sciences (BES) under Award DE-SC0021064. E.R., R.Z. and K.S.B performed part of his work while at the Kavli Institute of Theoretical Physics (KITP) that is supported by the National Science Foundation Grant No. NSF PHY-1748958.
\noindent \textbf{Author contributions:} K.S.B. conceived and supervised the project; W.L., G.N. and K.S.B. designed and conducted the differential conductance experiments, with the help of M.G., E.A., and K.F.; Exfoliation and AFM characterization was performed by W.L., G.N., K.H., M.G., V.L., J.V., V.B., and Q.M.; The h-BN/FTS devices were fabricated by G.N., W.L., and K.H.; C.F. and J.X. conducted the Kerr experiments and analysis; W.L. and G.N. analyzed the differential conductance data with help from W.K.P.; The theoretical discussion is contributed by R.Z. and E.R.; E.R., R.Z., and X.G. performed the calculations; G.G. provided the FeTeSe materials; T.T. and K.W. provided the h-BN materials; J.H. and J.C. performed the STEM measurement and EDS mapping; Q.T. and X.L. performed the EDS measurement; H.Z. and A.D. performed the ADF-STEM measurement; W.L. and G.N. drafted the manuscript with the help of J.X. and K.S.B.; All authors contributed to the discussion of the manuscript. 

\noindent \textbf{Data and materials availability:} All data are available in the manuscript or the supplementary materials.

\newpage
\section*{Figure Legend}

\begin{figure}
    \centering
    \includegraphics[width=1\linewidth]{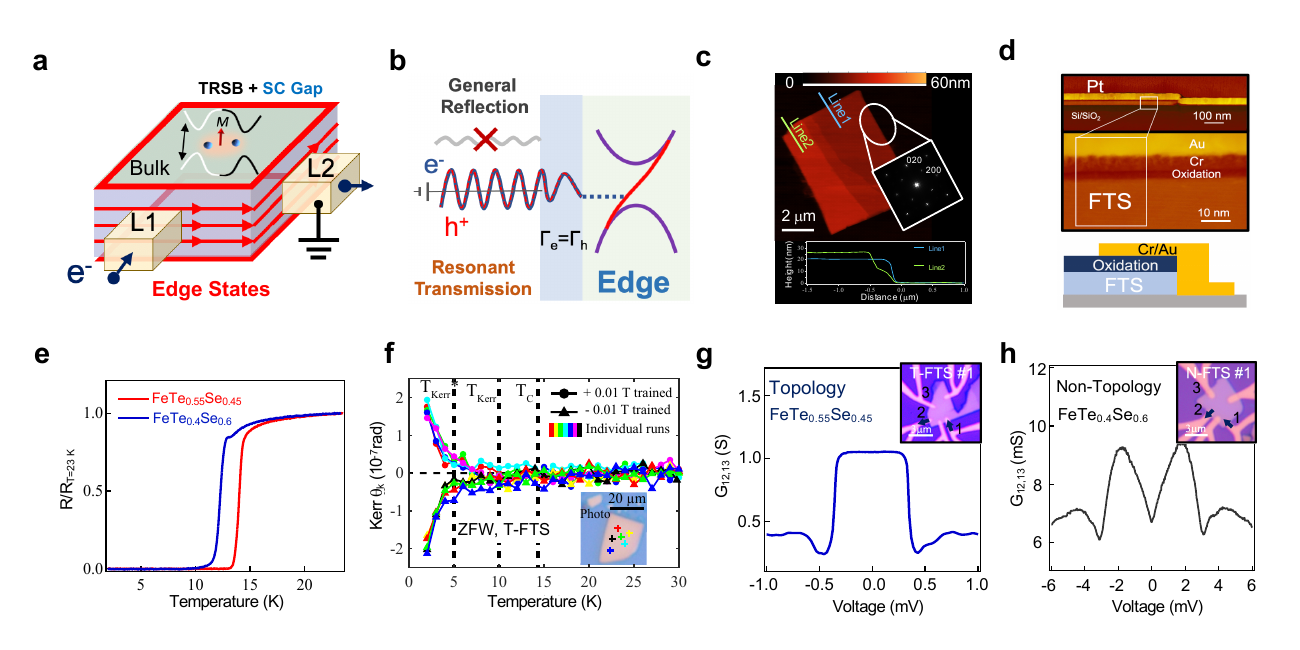}
    \label{Fig. 1}
\end{figure}

\noindent\textbf{Fig. 1. Non-local charge transport mediated by TSC edge states.} \textbf{a,} Magnetic order in a topological superconductor. When TRS is broken in the bulk of superconductor with topological band structures, the spontaneous fields emerge in the bulk and producing C-TSC with propagating edge modes along the side surface. Here, the gapless edge modes can mediate long-range non-local transport. \textbf{b,} The self-hermitian property of TSC edge modes prevents normal reflections. \textbf{c,} AFM and TEM electron diffraction pattern confirm that edges are sharp and cleaved along crystalline directions [100] or [010]. The lines in the topography image indicate the line profile locations. \textbf{d,} Cross-sectional ADF-STEM images of an \ch{FeTe_{0.55}Se_{0.45}} device. The top-surface part of the electrode is floated by remaining oxidized layer, which promise a primarily edge contact. \textbf{e,} The temperature dependence of the 4-terminal resistance of trivial-phase \ch{FeTe_{0.4}Se_{0.6}} and topological-phase \ch{FeTe_{0.55}Se_{0.45}} devices, normalized by the resistance at 23 K. \textbf{f,} SMOKE measurements during zero-magnetic-field warming on an Fe(Te, Se) flake after cooling in $+$0.01 T (circles) and $-$0.01 T (triangles) training fields. Different colors represent measurements at various locations. Inset: photo of the Fe(Te,Se) flake, with measurement locations marked with ``+''. \textbf{g,} Differential conductance ($G_{12,13}$ or $ G_{DE}$) versus bias-voltage measurements taken at 1.4 K on a topological \ch{FeTe_{0.55}Se_{0.45}} device (T-FTS\#1). Inset: the device image, arrows indicate the source/drain contacts and current direction. \textbf{h,} Differential conductance measured in the same configuration as \textbf{g,} but on trivial-phase \ch{FeTe_{0.4}Se_{0.6}} (N-FTS\#1). Inset: the device image.

\newpage
\begin{figure}
    \centering
    \includegraphics[width=1\linewidth]{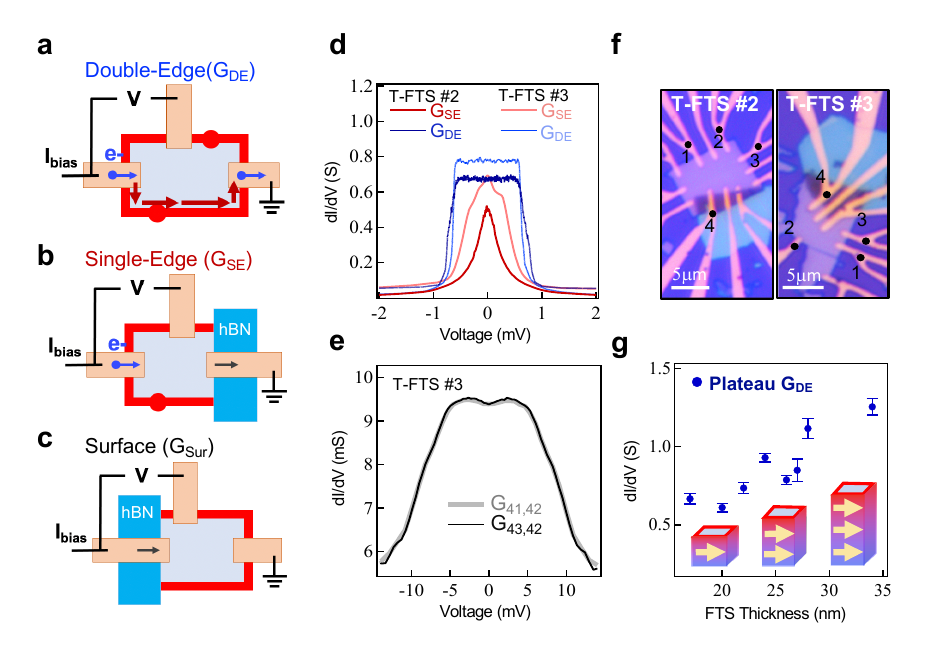}
    \label{Fig. 2}\vspace{-4ex}
\end{figure}

\noindent\textbf{Fig. 2. Configuration dependent conductance spectra in Topological Fe(Te,Se).} \textbf{a-c,} Schematics of the lead locations in the double-edge-lead ($G_{DE}$), single-edge-lead ($G_{SE}$), and surface -lead ($G_{Sur}$) configurations. \textbf{d,} Corresponding G-V curves in devices T-FTS\#2 and \#3 probed in single-edge and double-edge-lead configurations at 1.4 K and under zero magnetic field. \textbf{e,} G-V curves in device T-FTS\#3 probed in the surface-lead configuration, exhibiting spectra typical for normal AR\cite{park2010strong,tang2019quasi}. \textbf{f,} Images and lead number of T-FTS\#2 and \#3, where flakes are partly covered by an h-BN film (cyan) to allow purely surface contact. \textbf{g,} The magnitude of $G_{DE}$ at zero-bias voltage and base temperature versus the thickness of flakes in the double-edge-lead configuration. Inset: Schematic diagram illustrates that the flake thickness affects the number of channels allowing charge transport through the edge states.

\newpage
\begin{figure}
    \centering
    \includegraphics[width=0.75\linewidth]{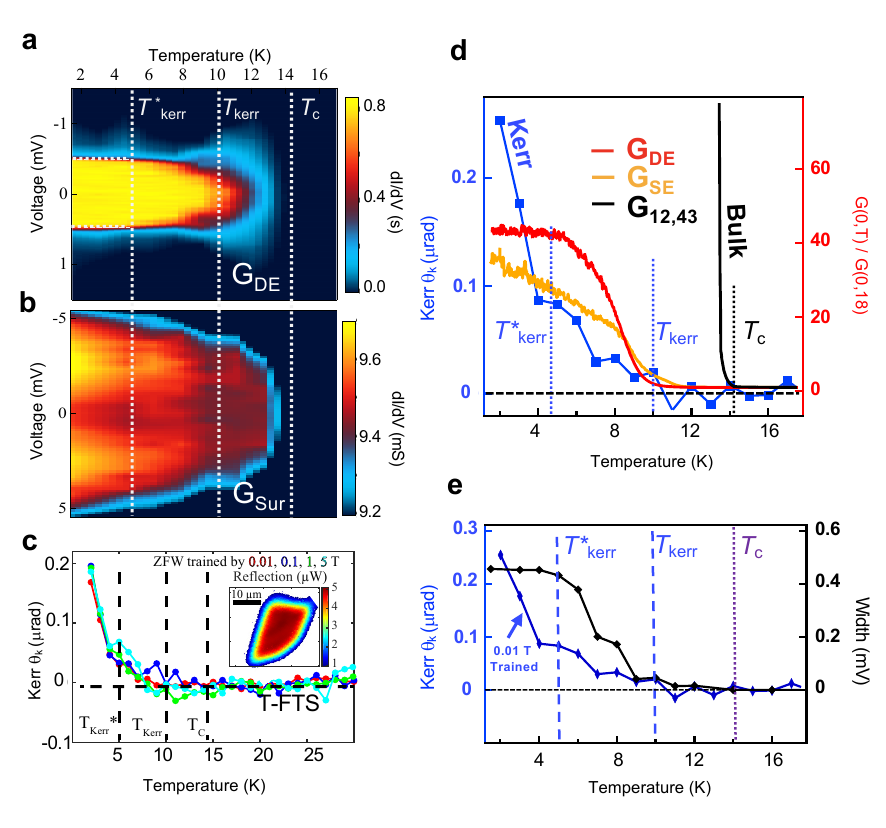}
    \label{Fig. 3}
\end{figure}
\noindent\textbf{Fig. 3. Temperature evolution.} 
\textbf{a,} False-colour map of the $G_{DE}$ temperature evolution vs V spectra for the non-local conductance plateau. Three regimes are observed: $G_{DE}$ is temperature independent for $ T < 5 K$, the plateau width slowly shrinks for $5 K < T < 10 K$, and the remaining features diminish rapidly above 10 K. \textbf{b,} False-colour map of the $G_{Sur}$ temperature evolution vs V spectra, consistent with local AR process. \textbf{c,} The $\theta_{k}$ curve for a \ch{FeTe_{0.55}Se_{0.45}} device, warmed in zero applied field using various training fields. A magnetic-order signal emerges below $T_{Kerr} \approx10~K$ , grows rapidly below $T^{*}_{Kerr} \cong5~K$. Inset: reflection image. \textbf{d,} The temperature evolution of the polar-Kerr $\theta_{k}$ and normalized differential conductance at zero-bias voltage $[G(0,T)/G(0,18~K)]$. Here, $G_{DE}$ and $G_{SE}$ correspondingly show the temperature evolution of the plateau and ZBCP cases in a 3-terminal configuration. $G_{12,43}$ is probed in a 4-terminal configuration (as indicated in Fig. 2f) to extract the temperature dependence of bulk super-current and ${T}_{c}$. \textbf{e,} Comparison of the temperature evolution of the polar-Kerr $\theta_{k}$ and plateau width from \textbf{a,}, both are measured on the same \ch{FeTe_{0.55}Se_{0.45}} flake.

\newpage
\begin{figure}
    \centering
    \includegraphics[width=0.75\linewidth]{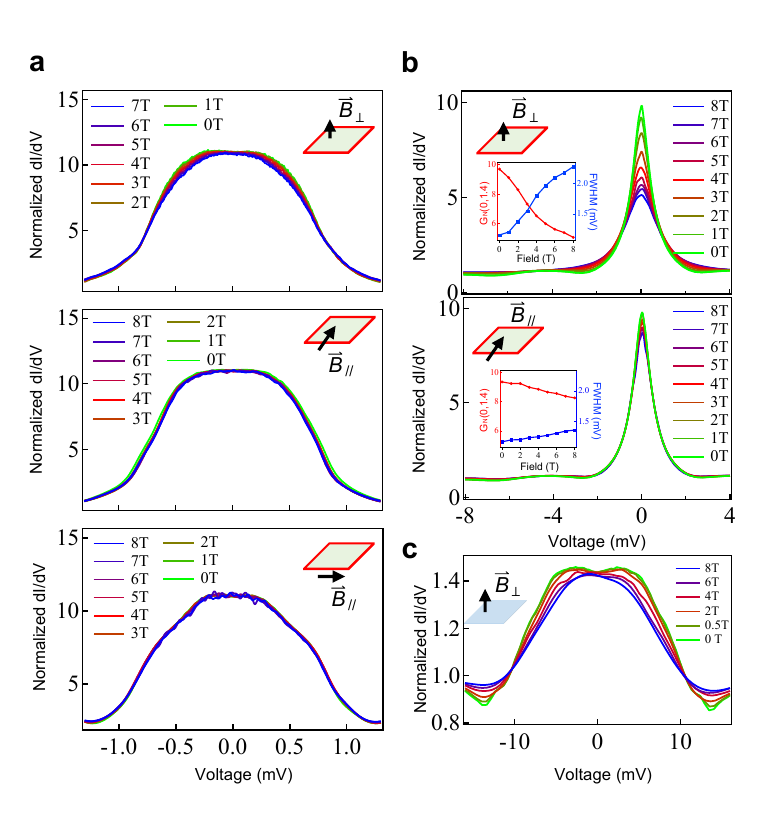}
    \label{Fig. 4}
\end{figure}

\noindent\textbf{Fig. 4. Response to external magnetic field.} \textbf{a,} $G_{DE}$ measurements taken with a magnetic field applied along different directions of the \ch{FeTe_{0.55}Se_{0.45}} flake at 1.4 K. Going from the upper to lower panels, the direction of the applied magnetic field is aligned out-of-plane, in-plane perpendicular, and in-plane parallel to the current into the edge of the flake, respectively. A second source on the edge perpendicular to the first provided the current parallel to the magnetic field. All curves are normalized by the relatively high-bias value. \textbf{b,} $G_{SE}$ measurements in an applied magnetic field. Top panel: magnetic field is aligned out-of-plane. Inset: Field dependence of the ZBCP height and full width at half maximum (FWHM). Bottom panel: magnetic field is aligned in-plane. \textbf{c,} $G_{Sur}$ measurements under the out-of-plane magnetic field. The suppression with an out-of-plane field in \textbf{b,} and \textbf{c,} is consistent with the local processes suppressed via the Doppler\cite{tanaka2003doppler} and vortex-state smearing effects\cite{naidyuk1996magnetic}. 

\section*{Competing interests} 
The authors declare no competing interest. 

\noindent\textbf{Supplementary Information}

\end{document}